\documentclass{article}
\usepackage{arxiv}
\usepackage{authblk}
\usepackage{graphicx}
\usepackage{float}
\usepackage{amsmath}
\usepackage[backend=bibtex,sorting=none,style=nature]{biblatex}
\usepackage{setspace}
\title{Exploration of Hexagonal, Layered Carbides and Nitrides as Ultra-High Temperature Ceramics}
\author[1]{Kat Nykiel}
\author[2]{Brian Wyatt}  
\author[2]{Babak Anasori}
\author[1]{Alejandro Strachan}

\affil[1]{School of Materials Engineering and Birck Nanotechnology Center, Purdue University}
\affil[2]{School of Materials Engineering, Purdue University}

\bibliography{zotero.bib}
\date{\today}

\begin{document}

\maketitle

\begin{abstract}

    Layered, hexagonal crystal structures, like zeta and eta phases, play an important role in ultra-high temperature ceramics, often significantly increasing toughness of carbide composites. Despite their importance open questions remain about their structure, stability, and compositional pervasiveness. We use high-throughput density functional theory to characterize the thermodynamic stability and elastic constants of layered carbides and nitrides M$_{n+1}$X$_{n}$ with $n$ = 1, 2, and 3, $M$ = Ta, Ti, Hf, Zr, Nb, Mo, V, W, Sc, Cr, Mn  and $X$ = C, N. The stacking sequences explored are inspired by the possible use of MXenes as precursors to enable relatively low temperature processing of high-temperature ceramics. We identified 67 new hexagonal, layered materials with thermal stability comparable or better than previously observed zeta phases. To assess their potential for high temperature applications, we used machine learning and physics-based models with DFT inputs to predict their melting temperatures and discovered several candidates on par with the current state of the art zeta-like phases and five with predicted melting temperatures above 2500 K. The findings expand the range of chemistries and structures for high-temperature applications.
    
\end{abstract}

Materials that can perform at extreme temperatures and other stressors are of interest in hypersonics, nuclear reactors, and even solar power plants \cite{leeNuclearApplicationsUltraHigh2014, squireMaterialPropertyRequirements2010, silvestroniOverviewUltrarefractoryCeramics2019}. Ultra-high temperature ceramics (UHTCs), typically defined as materials with stable properties above 2300 K, include the highest-known melting temperature materials, found in the Hf-C system at $\sim$4500 K \cite{cedillos-barrazaInvestigatingHighestMelting2016}. The cubic TaC phase, another Group VB carbide, achieves a comparably high melting temperature of $\sim$4200 K \cite{cedillos-barrazaInvestigatingHighestMelting2016}, while the hexagonal Ta$_2$C phase has a melting temperature of $\sim$3600 K \cite{morrisMicrostructuralFormationsPhase2012}. Notably, the Ta-C system forms an additional phase, the hexagonal $\zeta$-Ta$_4$C$_{3-x}$ phase, which is advantageous due to its high reported fracture toughness of 15.6 ± 0.5 MPa m$^{1/2}$ and fracture strength of 508 ± 97 MPa \cite{sygnatowiczProcessingCharacterizationZTa4C}. This high fracture toughness has been attributed to cleaving along the missing carbon layers in the vacancy-ordered $\zeta$-Ta$_4$C$_{3-x}$ phase \cite{sygnatowiczProcessingCharacterizationZTa4C}. This Ta$_4$C$_{3-x}$ $\zeta$-phase has a reported upper stability limit of $\sim$2400 K \cite{schwindThermalPropertiesElastic2020}, though others have suggested this phase is metastable up to 3200 K \cite{gusevAtomicVacancyOrdering2007}. In addition, the Ta$_3$C$_2$ phase has been suggested as the perfectly ordered zeta phase in the Ta-C system, with the Ta$_4$C$_{3-x}$ phase being metastable \cite{weinbergerComputationalSearchZeta2019}. Non-Ta $\zeta$-M$_4$C$_{3-x}$ carbides has been experimentally observed in V-C and Nb-C multi-component systems \cite{wiesenbergerReactiveDiffusionPhase1998}, though disagreements about their upper stability limits exist \cite{schwindThermalMechanicalProperties2021}. Reports of these limits range from 1590 K \cite{ghaneyaInvestigationVanadiumSolid1985} to 1980 K \cite{wiesenbergerReactiveDiffusionPhase1998} for V-C and 1850 K \cite{wiesenbergerReactiveDiffusionPhase1998} for Nb-C. The $\zeta$-Nb$_4$C$_{3-x}$ phase exhibits slow formation kinetics \cite{wiesenbergerReactiveDiffusionPhase1998}, making its upper stability limit comparatively difficult to study. Additional work is needed to clarify the upper stability limits of these phases. This study explores the thermodynamical stability and melting of a wide range of possible layered, hexagonal structure and chemical space. 

One factor complicating the determination of stability of zeta and similar phases is the high temperature and pressure required for their formation. The synthesis of $\zeta$-Ta$_4$C$_{3-x}$ is performed by pressing powders at at 2100 K \cite{schwindThermalPropertiesElastic2020,hackettPhaseConstitutionMechanical2009} or by pressureless sintering at 2270-2300 K \cite{gusevAtomicVacancyOrdering2007}. Alternative synthesis routes are therefore highly desirable. In this work, we explore layered hexagonal phases that could be produced by stacking 2D MXenes. This process can provide a lower-temperature processing alternative to high-temperature ceramics. Stacking of 2D materials to form bulk structures has been documented in other chemical spaces, with successful synthesis reported in transition metal dichalcogenides \cite{zhangPhononRamanScattering2015} and BaFZnP \cite{carioDesigningNewInorganic2005}, among other systems. MXenes, a class of layered 2D materials where $M$ denotes a transition metal and $X$ is carbon or nitrogen, are a promising alternative as recently proposed \cite{wyattUltrahighTemperatureCeramics2023}. MXenes contain alternating $M$ and $X$ layers in an FCC sequence in the ratio $M_{n+1}X_n$, for $n$=$1,2,3$ and share their chemical space with UHTCs. The five transition metals commonly used in UHTCs have been successfully incorporated into MXenes: Hf \cite{zhouSynthesisElectrochemicalProperties2017}, Ta \cite{linTheranostic2DTantalum2018}, Zr \cite{zhouTwoDimensionalZirconiumCarbide2016}, Ti \cite{naguibTwoDimensionalNanocrystalsProduced2011}, and Nb \cite{naguibNewTwoDimensionalNiobium2013}. Stacking of 2D MXene sheets into bulk systems creates a set of vacancy-ordered \textit{zeta-like}, layered structures; for example, sintering Ti$_3$C$_2$T$_X$ sheets has resulted bulk layered structures \cite{wyattHightemperatureStabilityPhase2021}. We systematically explore the stability, structure, and elastic constants of the 352 layered bulk structures that can be formed by stacking up to three $M_2X$, $M_3X_2$, or $M_4X_3$ layers. We also expand the chemical space beyond Group VB elements with $M$ = Ta, Ti, Hf, Zr, Nb, Mo, V, W, Sc, Cr, Mn and $X$ = C, N. We discovered 67 new materials with stability comparable or better than the ones previously reported. We use ML and physics-based models to explore the thermal stability of these alloys and find several potential structures for high-temperature applications.

\section*{Results}

\begin{figure}
    \centering
    \includegraphics[width=\textwidth]{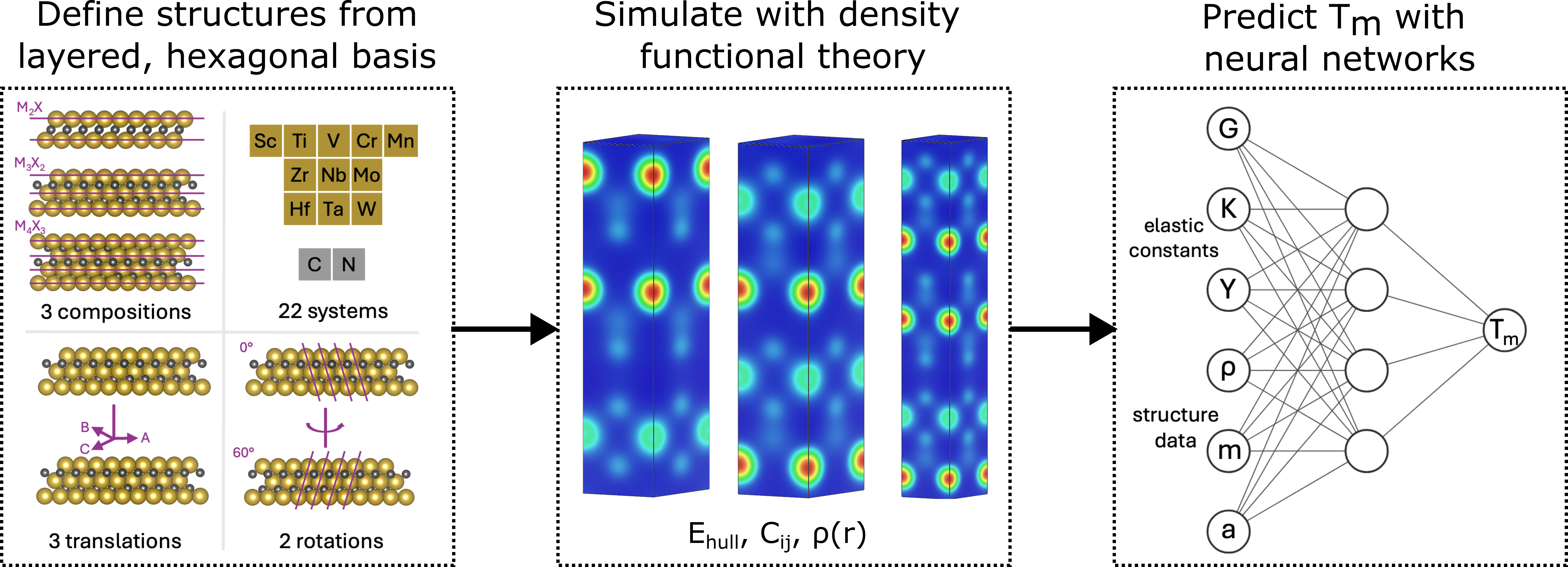}
    \caption{Workflow schematic. The domain of investigated hexagonal, layered structures includes is defined in the first panel. The second panel shows several of the unique unit cells, on which we run density functional theory. We use the DFT results as inputs to a parsimonious neural network, as shown in the third panel.}
    \label{fig:workflow-schematic}
\end{figure}

\begin{figure}[H]
    \centering
    \includegraphics[width=\textwidth]{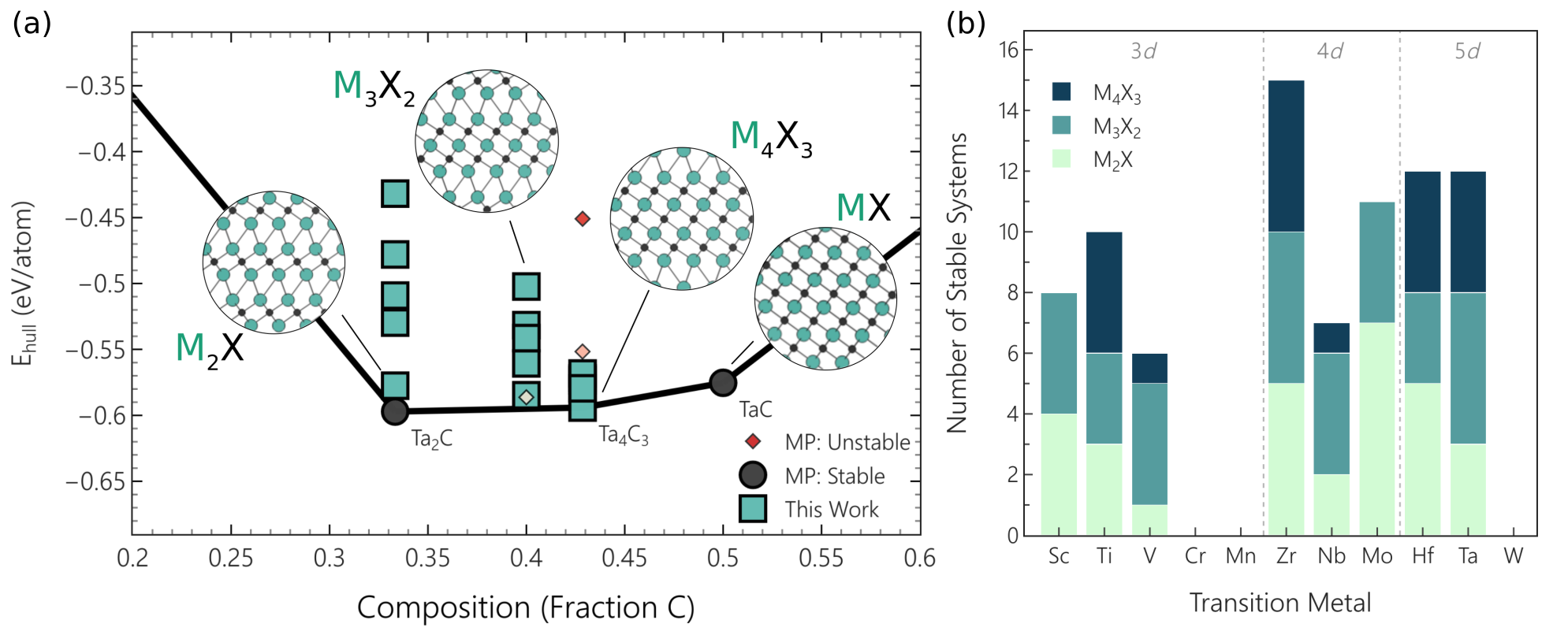}
    \caption{(a) Convex hull for Ta-C at the zeta-phase composition range and (b) the number of mechanically stable structures within 0.1 eV/atom of the convex hull for each chemical system.}
    \label{fig:convex-hull}
\end{figure}

Figure \ref{fig:workflow-schematic} summarizes our approach. We start with the 16 unique layered unit cells resulting from stacking MXene-like sheets, including rotations and translations that change the stacking sequence. These 16 structures are listed in the Supplemental Information. We describe the stacking sequence of each structure using the letters $h$ and $c$ to indicate the hexagonal (ABA) or cubic (ABC) stacking around the layer, following Jagodzinski-Wyckoff notation. M-M layers are indicated by a dash in the sequence, e.g. $hch$-$hch$. An example of this descriptor is provided in Figure \ref{fig:structure_generation} in Methods for the prototypical vacanacy-ordered zeta phase. We refer to these structures as vacancy-ordered because the M-M layers between sheets contain an ordered plane of carbon vacancies. Considering the chemical space of 11 metals, C and N, we used DFT to relax and compute the elastic constants of 352 materials, see Methods for the simulation details. We computed the thermodynamic stability of all configurations by constructing the convex hull of each chemical system, including all available structures within each chemical system from the Materials Project \cite{jainCommentaryMaterialsProject2013}. Figure \ref{fig:convex-hull} shows the convex hull for Ta-C, the prototypical UHTC system, with insets demonstrating possible M-X layering at each composition, and the number of stable systems for each metal system. Interestingly, we find that the majority of layered carbides have systems on or near the convex hull, while the nitride systems lie higher in energy. The Hf, Mo, Nb, Ta, and Zr carbide systems have structures on the convex hull, with all three layered compositions of zeta-like phases on the convex hull for the Nb-C and Zr-C systems. This demonstrates that, with respect to the competing phases in the Nb-C and Zr-C systems, the $hcch$-$hcch$-$hcch$ zeta phases are the most energetically stable. The Ta-C system has one structure on the convex hull: the $hcch$-$hcch$-$hcch$ Ta$_4$C$_3$ phase, We note that the majority of the zeta phase nitrides are less stable.

On average, structures obtained by stacking $M_2X$ layers have lower energies above the convex hull than those derived from $M_3X_2$ and $M_4X_3$ layers. The average energy above hull for $M_2X$ is 0.15 $\pm$ 0.11 eV/atom, compared to 0.18 $\pm$ 0.13 eV/atom for $M_3X_2$ and 0.21 $\pm$ 0.14 eV/atom for $M_4X_3$. The higher concentration of $M$-$M$ layers in the $M_2X$ phases results in a lower energy above hull despite the $M$-$M$ layers expected to be less energetically favorable than then $M$-$X$ layers. This is likely because the $M_4X_3$ with less $M$-$M$ vacancies must compete against more stable, cubic, $MX$ stoichiometric phases with no $M$-$M$ layers. We provide a table of the average properties of each stacking sequence in the SI. 

\begin{figure}[H]
    \centering
    \includegraphics[width=0.5\textwidth]{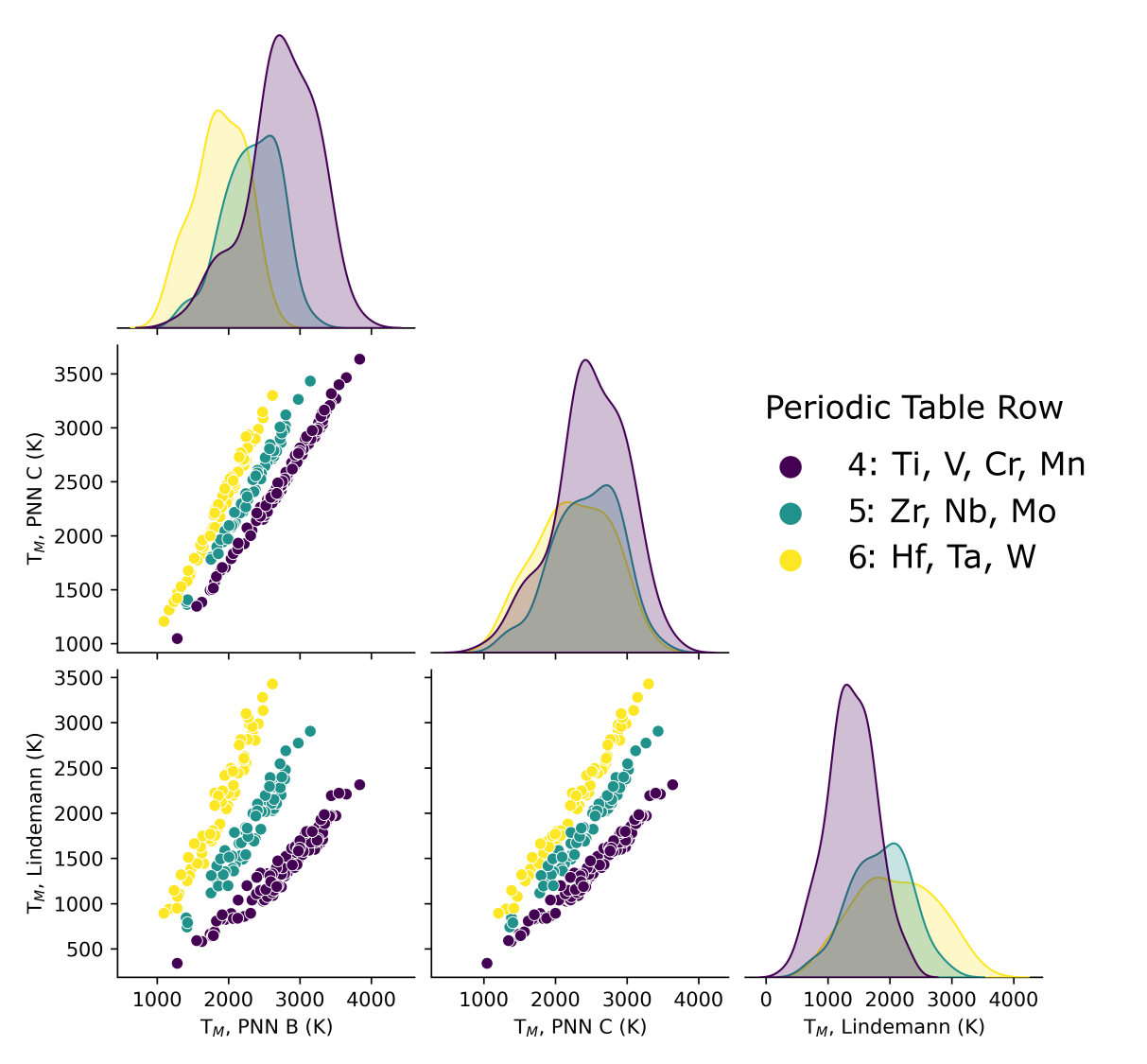}
    \caption{Predicted melting temperatures using PNNs B and C, and Lindemann melting criterion.}
    \label{fig:melting-temp-predictions}
\end{figure}

We now assess the potential of these structures for high-temperature applications by estimating their melting temperatures. The direct calculation of melting temperatures from DFT calculations is computationally intensive \cite{mishraComparingAccuracyMelting2022}, thus we use elastic constants as inputs to expressions derived machine learning and physics to estimate melting temperatures. As described in the Methods section, we employ both parsimonious neural networks \cite{desaiParsimoniousNeuralNetworks2021} and Lindemann melting law. The supplemental information provides the expressions written as a function of four scaled variables with unit of temperature associated with the Debye temperature, characteristic atomic spacing, and bulk and shear moduli. Figure \ref{fig:melting-temp-predictions} shows the predicted melting temperatures for all investigated zeta-like phases and correlations between the various models. The distributions of predicted melting temperatures for three models show a large number of materials with melting temperatures above 2500 K. The off-diagonal panels in Fig. \ref{fig:melting-temp-predictions} show strong correlations between the various models for melting temperature. We observe the clustering of predictions between the three models which originates from periodic table row of the $M$ element in the structure. Although the models are trained across the periodic table, their dependence on the atomic mass differs. The inclusion of elements from three periodic table rows leads to a tri-modal distribution; PNN B has no explicit atomic mass input, while PNN C and Lindemann both include differing expressions including mass, as discussed in the Supplemental Information.

\begin{figure}[H]
    \centering
    \includegraphics[width=\textwidth]{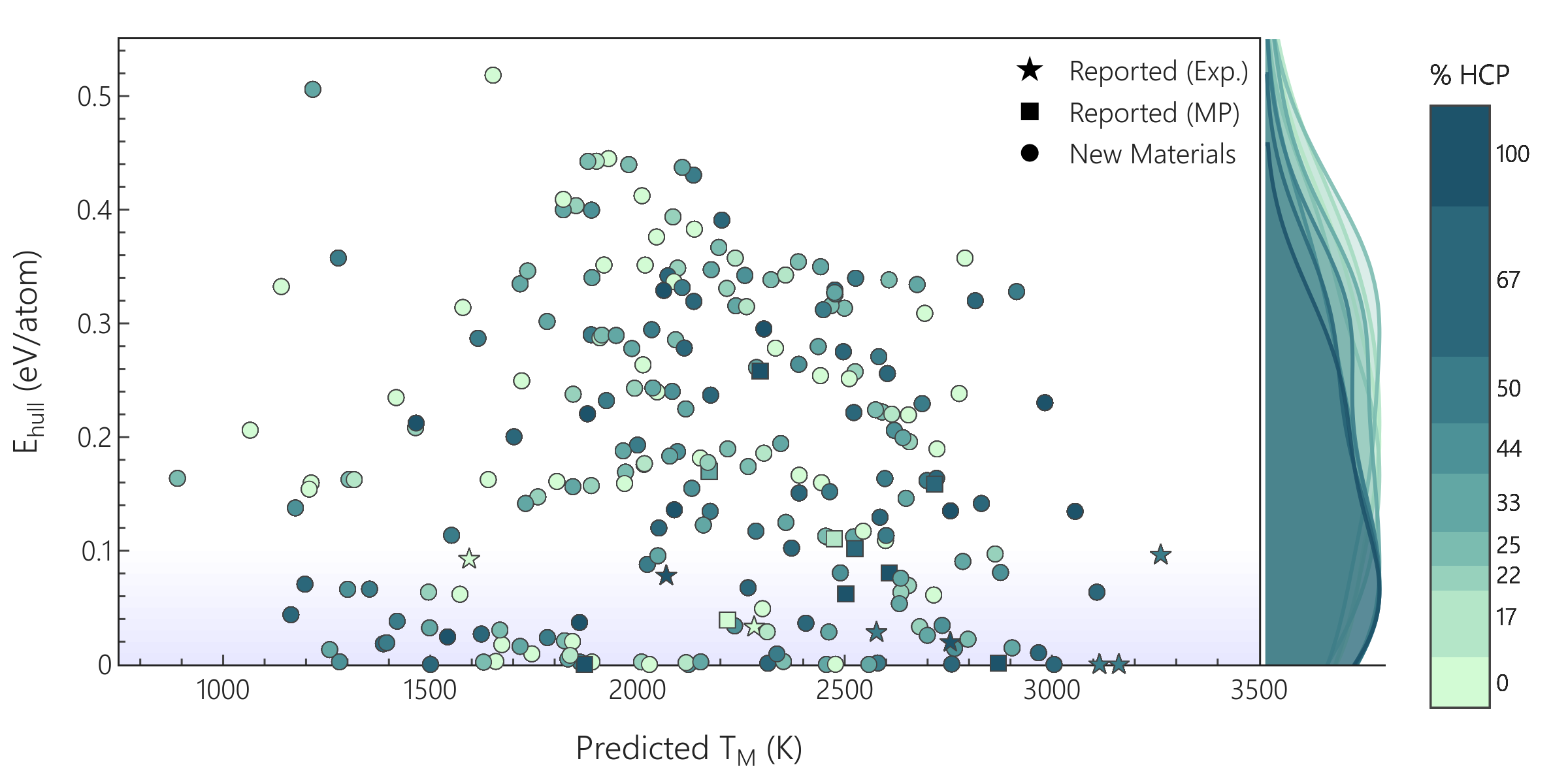}
    \caption{Distribution of zeta-like phases, where color indicates the fraction of HCP stacking in the HCP/FCC stacking sequence.}
    \label{fig:uhtc-distribution}
\end{figure}

To provide a single estimate of the melting temperature for each system, we use the average of the three prediction methods. We find the $M_2X$, $M_3X_2$, and $M_4X_3$ phases have average melting temperatures of 2060, 2200, and 2200 K, respectively. This is likely due to the higher concentration of $M$-$M$ vacancies in the $M_2X$ phases, as strong ionic/covalent $M$-$X$ bonds generally yield higher melting temperatures than metallic $M$-$M$ bonds \cite{wyattUltrahighTemperatureCeramics2023}. In addition, we find that, on average, HCP and mixed HCP/FCC stacking sequences result in higher melting temperature predictions, with averages of 2300 and 2200 K, while FCC stacking sequences have an average of 2000 K. 
Figure \ref{fig:uhtc-distribution} compares the stability and estimated melting temperatures for all the systems explored. Circles represent structures reported for the first time here. Our systematic exploration yielded a total of 81 structures within 0.1 eV/atom of the convex hull. These include all the experimentally observed phases (stars) providing support for the use of energy above convex hull as a of synthesizability. 58 of these structures lie within 0.05 eV/atom of the convex hull and 9 lie on the convex hull. These nine structures are predominately populated by the $hcch$-$hcch$-$hcch$ and $hch$-$hch$-$hch$ structures for $n=3$ and $n=2$, respectively. We note that, in general, the HCP stacking sequence is more stable than the FCC stacking sequence, with the HCP stacking sequence having a lower average energy above the convex hull, as demonstrated by the violin plots on the right side of Figure \ref{fig:uhtc-distribution}.

Figure \ref{fig:uhtc-distribution} reveals a range of new structures of interest for high-temperature applications. The three highest melting-temperature systems (V$_4$C$_3$, Nb$_4$C$_3$, and Ta$_4$C$_3$) all exhibit the $hcch$-$hcch$-$hcch$ stacking sequence and have been experimentally reported. We identified several new phases which lie on the convex hulls of their chemical systems and are predicted to have melting temperatures above 2500 K. The include Nb$_3$C$_2$ ($hch$-$hch$-$hch$), Mo$_2$C ($hh$-$hh$-$hh$, $hh$-$hc$-$hc$, $ch$-$cc$-$hc$), and Zr$_4$C$_3$ ($hcch$-$hcch$-$hcch$). In addition to these systems, we identified systems that lie within 0.1 eV/atom from the convex hull, with melting temperatures above 2900 K: V$_3$C$_2$ ($hch$-$hch$-$hch$), Sc$_3$C$_2$ ($ccc$-$ccc$), and Ta$_4$C$_3$ ($hcch$-$cccc$-$hcch$). These systems are promising candidates for future experimental investigation as UHTCs.

\section*{Discussion}

To assess the accuracy of our predictions, we investigate the three $\zeta$-$MX_{0.67}$ phases reported experimentally: the $\zeta$-Ta$_4$C$_3$, $\zeta$-V$_4$C$_3$, and $\zeta$-Nb$_4$C$_3$ systems. For Ta$_4$C$_3$ zeta phase we predict a melting temperature of 3110 K, this is more consistent with reports by Gusev et al. that this material is stable up to 3200 K \cite{gusevAtomicVacancyOrdering2007} than the value of $\sim$2400 K reported in other literature \cite{schwindThermalPropertiesElastic2020}. The predicted Young's and shear moduli for this phase, 180 and 460 GPa, respectively, are in good agreement with the experimental values of Schwind et al. \cite{schwindThermalPropertiesElastic2020} who reported  149.8 GPa and 379 $\pm$ 5 GPa. 
For the $\zeta$-V$_4$C$_3$ and $\zeta$-Nb$_4$C$_3$ phases, we predict melting temperatures of 3500 K and 3400 K, respectively, the highest among the zeta-like phases we investigated. Their experimental upper stability limits range from 1600 K \cite{ghaneyaInvestigationVanadiumSolid1985} to 1980 K \cite{wiesenbergerReactiveDiffusionPhase1998} for V-C and 1850 K \cite{wiesenbergerReactiveDiffusionPhase1998} for Nb-C. This suggests that rather than melting, the reported upper stability limit of these phases is due to their decomposition into other phases. We find the $\zeta$-V$_4$C$_3$ lies off the convex hull by 0.1 eV/atom, while the $\zeta$-Nb$_4$C$_3$ phase lies on the convex hull. This could provide an explanation for the reported difficulty in synthesizing the $\zeta$-V$_4$C$_3$ phase.

In conclusion, we characterized the stability and elasticity of a wide range of layered, hexagonal carbides and nitrides using DFT. To assess the potential of these phases for high temperature applications we use the DFT calculations with physics and machine learning expressions to estimate their melting temperatures. We report several new stacked hexagonal, layered carbides which lie on the convex hull, particularly in the Ta-C, Nb-C, and Zr-C systems. Among these zeta-like structures on the convex hull, we identify five new structures with melting temperatures above 2500 K in the Nb-C, Mo-C, and Zr-C chemical systems.

\section*{Methods}

\subsection*{Structure Generation}

\begin{figure}[H]
    \centering
    \includegraphics[width=0.8\textwidth]{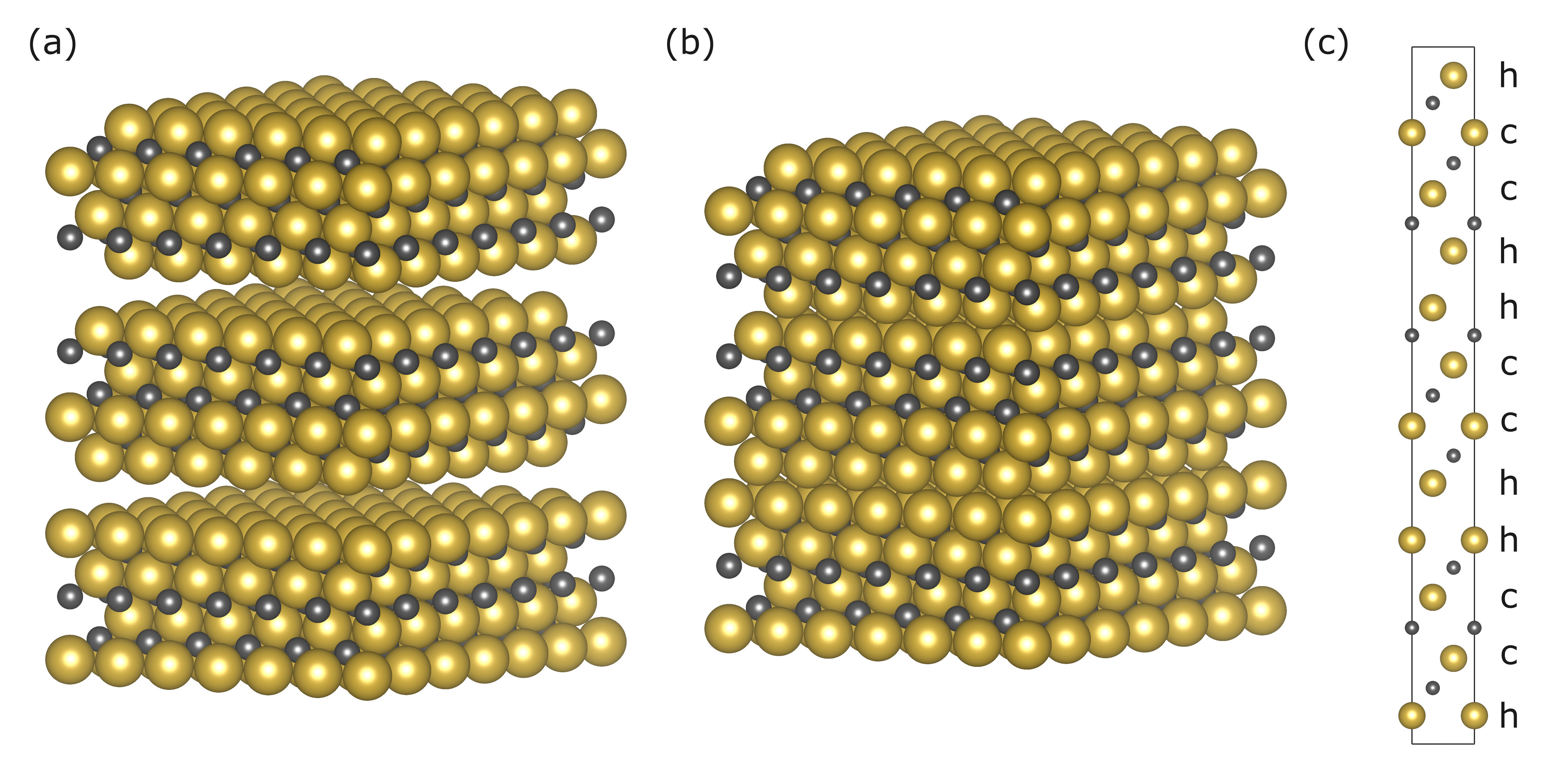}
    \caption{(a) Delaminated MXene flakes (b) Stacked hexagonal, layered carbide, forming a $\zeta$-phase structure (c) $\zeta$-phase unit cell}
    \label{fig:structure_generation}
\end{figure}

Previous computational predictions of zeta phases have been limited to the Group VB carbides \cite{weinbergerCrystalStructurePhase2019} and Hf-C-N \cite{tangDensityFunctionalTheory2024} systems. For our exploration of this stacked MXene domain, we use the full domain of mono-M MXene phases as the basic building blocks of our structures. We consider $M_2X$, $M_3X_2$ and $M_4X_3$ phases, with a composition space of $M$ = Ta, Ti, Hf, Zr, Nb, Mo, V, W, Sc, Cr, Mn as the $M$ transition metal and $X$ = C, N. In this work, we refer to the stacking sequence using the Jagodzinski-Wyckoff labeling system, with $c$ for any FCC stacking (ABC) and $h$ for any HCP (ABA) planes. The initial reporting of the zeta phase used this notation \cite{partheCrystalChemistryClose1970}. This compact notation is useful to describe the properties of our zeta-like phases and their relation to stacking sequence. A prototypical example of the transformation from MXene flakes to a $\zeta$-phase is shown Figure \ref{fig:structure_generation} with the zeta-phase structure in Fig \ref{fig:structure_generation}(c) as $hcch$-$hcch$-$hcch$.

To generate a complete list of possible stacking sequences and interfacial combinations of hexagonal, layered carbides and nitrides, we consider unit cells of up to 3 MXene-like layers stacked. We also consider two possible in-plane rotations of MXenes: zero and 60 degrees. We generated all of these combinations, then removed rotationally and translationally equivalent structures, and finally removed structures where $M$ atoms at $M$-$M$ interfaces align along the $a$ and $b$ axes in an unstable equilibrium. After this procedure, we identify 16 unique structures in total between the $M_2X$, $M_3X_2$, and $M_4X_3$ phases. We provide a complete list of these structures and their stacking sequences in the Supplemental Information. These structures resemble vacancy-ordered zeta phases, but span a larger space of interfacial combinations. We will collectively refer to these structures as zeta-like phases. This domain of zeta-like systems spans 4 space groups, with international space groups numbers of 186 (P6$_3$mc), 166 (R$\bar{\text{3}}$m), 164 (P$\bar{\text{3}}$m1), and 156 (P3m1), with 164 being the most common space group. The space group of the prototypical vacancy-ordered zeta phase in \ref{fig:structure_generation}(c) is 166 (R$\bar{\text{3}}$m).

\subsection*{Thermodynamic Stability}

We used density functional theory (DFT) to relax all the structures and predict their thermodynamic stability. We employed the Vienna \textit{ab initio} Simulation Package (VASP) \cite{kresseEfficientIterativeSchemes1996, kresseEfficiencyAbinitioTotal1996} with the Perdew-Burke-Ernzerhof (PBE) exchange-correlation functional \cite{perdewGeneralizedGradientApproximation1996} and projector augmented wave (PAW) pseudopotentials \cite{blochlProjectorAugmentedwaveMethod1994}. For geometry optimization, we used a plane wave kinetic energy cutoff of 550 eV and $1450 / $n$_{atoms}$ as the total number of k-points, dispersed evenly along the reciprocal lattice vectors \cite{ongPythonMaterialsGenomics2013}. We considered spin polarization, with initial magnetic moments of zero. We performed the calculations using the \textit{atomate2} library \cite{ganoseMaterialsprojectAtomate2V02024} to ensure reproducibility. To determine the thermodynamic stability of the zeta-like phases, we constructed the convex hull for each $M$-$X$ chemical system, for a total of 22 convex hulls, each with 16 zeta-like structures. We constructed each convex hull using all available Materials Project \cite{jainCommentaryMaterialsProject2013} structures within the $M$-$X$ chemical system. 

\subsection*{Elastic Constants}

In addition to thermodynamic stability, we also calculated the elastic tensors of the zeta-like phases to measure mechanical stability and estimate high-temperature elastic properties. This was done using the \textit{ElasticMaker} code of \textit{atomate2}, which implements the workflow introduced by de Jong et al. \cite{dejongChartingCompleteElastic2015}. We deformed each component of the strain tensor with strain magnitudes of $\delta \in \{-0.01,-0.005,+0.005,+0.01\}$. This creates a total of 24 deformations, before reducing this number due to symmetry. We relaxed each of these structures using the following DFT parameters: plane wave energy cutoff of 700 eV and k-point densities of 7000 per-reciprocal-atom (pra). We calculated the elastic tensor using the stress-strain relationship from each deformation following $\sigma_{ij} = C_{ijkl} \varepsilon_{kl}$. Cholesky decomposition \cite{choleskyNoteMethodeResolution1924} is used on each elastic tensor to ensure positive definiteness and positive eigenvalues to characterize the mechanical stability of each system. From the elastic tensor, we then derive the Voigt-Reuss-Hill bulk and shear moduli, Young's modulus, and Debye temperature. We validated our calculations with the overlap between our dataset and systems currently in Materials Project, and obtained RMSE values of 3.6 GPa for shear modulus, 1.7 GPa for bulk modulus, and 19 K for Debye temperature. More details of this comparison are provided in the Supplemental Information.

\subsection*{Melting Temperature Predictions}

To assess the high-temperature stability of our zeta-like phases, we estimated the melting temperature of all the MXene-derived structures. While molecular dynamics simulations \cite{meiFreeenergyCalculationsMelting1992} have been used to provide accurate estimates of the melting temperature, the computational intensity of the calculations preclude its use for the 352 predicted structures. Thus, we estimate the melting temperature from the fundamental DFT properties using the Lindemann melting law and similar expressions obtained using parsimonious neural networks (PNNs) \cite{desaiParsimoniousNeuralNetworks2021}. We provide the specific details of this implementation in the Supplemental Information. In this work, we use the two PNN models with the lowest RMSE error.

\subsection*{Acknowledgements}

We acknowledge support from National Science Foundation, Award 2124241. Thank you to Purdue's Rosen Center for Advanced Computing.

\subsection*{Author Contributions}

Kat Nykiel ran the DFT, performed the analysis, and wrote the manuscript. Brian Wyatt and Babak Anasori provided guidance chemical space to explore and directed the analysis. Alejandro Strachan provided guidance on the analysis and manuscript. All authors contributed to the manuscript.

\subsection*{Competing Interests}

All authors declare no financial or non-financial competing interests. 

\subsection*{Data Availability}

The VASP data and PNN inputs/outputs for this work is available on Zenodo at doi:10.5281/zenodo.16944619. 

\subsection*{Code Availability}

The code used to generate this workflow is available at https://github.com/katnykiel/mxene-nlc. 

\printbibliography

\end{document}